\documentclass[epj]{svjour}

\usepackage{graphics,graphicx}
\usepackage{amssymb,amsmath}
\usepackage{times,latexsym}

\begin{document}
\title{Renormalizing Sznajd model on complex networks taking into account the effects of growth mechanisms}
\author{M. C. Gonz\'alez\inst{1} \and 
        A. O. Sousa\inst{2} \and 
        H. J. Herrmann\inst{1,3}}
\institute{Institute for Computational Physics, 
           Universit\"at Stuttgart, Pfaffenwaldring 27, 
           D-70569 Stuttgart, Germany \and
           Systemgestaltung, ETH Zentrum
           Kreuzplatz 5, F 27.2
           8032 Zurich - Switzerland
           \and
           Departamento de F\'{\i}sica, Universidade Federal do
           Cear\'a, 60451-970 Fortaleza, Brazil}
\date{Received: date / Revised version: date}
\abstract{
We present a renormalization approach to solve the Sznajd opinion 
formation model on complex networks. For the case 
of two opinions, we present an expression of the
probability of reaching consensus for a given opinion as a function 
of the initial fraction of agents with that opinion. The calculations 
reproduce the sharp transition of the model on a fixed network, as
well as the recently observed smooth function for the model when 
simulated on a growing complex networks.
}
\maketitle

\section{Introduction}
Opinions can either be made up by a person or taken over from
another person. Sometimes some people try to force
their o\-pi\-nions on others. In general, all people are
free to form opinions as they see fit. The mechanism 
of opinion formation is ``normative'', i.e., normative
in the sense of what {\it ought to be}, opposed to a 
``positive'' mechanism, which is based on observation 
{\it what is}~\cite{wiki}. Based on this facts, and with 
the necessary simplifying assumptions, socio-physics 
gave the opportunity to apply techniques of statistical 
physics to model opinion formation among
people~\cite{stauffer,galamI,galamII}. 

One of the opinion formation
models that has generated immediate interest in many
authors on the field is the Sznajd model \cite{sznajd1}, which 
is based on the slogan ``together we stand'': Individuals are represented
by the lattice nodes (one-dimensional in its first version), 
and each randomly selected pair of neighbors convinces all their neighbors 
of their opinions, if and only if the pair shares the same opinion; 
otherwise, the neighbors' opinion are not affected. It differs from 
other consensus models by dealing only with communication between neighbors, 
and the information flows outward as in rumor spreading: a site does 
not follow what the neighbours tell the site. 

On networks with fixed size, the results of the model do not depend 
much on the spatial dimensionality and type of neighborhood selected 
(i.e., two nodes convince the others, three convince the others, etc.)
\cite{stauffer2,sousa,schulze}. In the case of $q$ choices of opinion, the model has
$q$ homogeneous absorbing states, where all individuals
choose the same opinion; in the context of opinion, one says
the system reaches consensus. The case of two opinions ($q=2$) has been
the most studied, denoting opinions as Ising variables ``up'' or $+1$, 
and ''down'' or $-1$. In more than one dimension, 
the probability ($P_{up}$) of reaching consensus 
``all up''  depends on the initial fraction $p$ of individuals with
opinion ''up''; for $p>0.5$, the probability of reaching ``all up'' 
as stationary state is close to one, while for $p<0.5$ it is 
negligible, having a sharp transition in $p=0.5$, which can be 
interpreted as a dynamical phase transition. Computer simulations 
in \cite{stauffer3} indicate that the universality class associated
with this dynamical phase transition is different from the
universality class of the Ising model. The distribution of time 
needed to reach the stationary state is a peak followed by a fast 
decay \cite{slanina}.

Much less is known about the Sznajd model on growing
networks. Interactions of groups of people in some circumstances 
can be thought as a growing system, i.e., in a city with 
positive rate of immigration. In a first and simple approximation, 
it can be modeled by a growing scale-free network~\cite{Barabasi:RMP}. 
Recently, applying a Sznajd model recipe not after the complete
network has been constructed, but while the network grows, i.e, while
each new node is added to the network, one could observe that the
Sznajd model simulated on scale-free networks, Barab\'asi-Albert
network and a pseudo-fractal network~\cite{Doro}, 
the system reaches consensus~\cite{bonnekoh,us}. But in contrast to the sharp
transition observed for the networks of fixed size, in which 
the Sznajd recipe is performed only after building up completely the 
network, the probability that the system reaches ``all up''  for a
growing complex networks is a smooth function of $p$. In addition, 
this function depends on the type of neighborhood selected.

In this work, we propose a real space renormalization approach 
\cite{Reynolds} to calculate the probability $P_{up}(p)$ of 
reaching consensus on opinion ``up'' as a function of the initial 
fraction $p$ of opinion ``up''. Our results are for two 
common rules of neighborhood, namely ``$r$-convince all their
neighbors'', with $r=2$ and $r=3$. We have obtained the 
two well-known results known for the model: a smooth function of $p$
for the growing case and an expression which approximates the 
step function for fixed networks.

In the next section, we present the hierarchical network
used in our calculations. Then, we present the renormalization
approach and the analytical expressions obtained, each case
is compared with the results from the numerical simulations, 
previously reported in \cite{us}, as well as for the $BA$ 
scale-free network. 

\begin{figure}[!hbt]
\includegraphics[width=8.5cm]{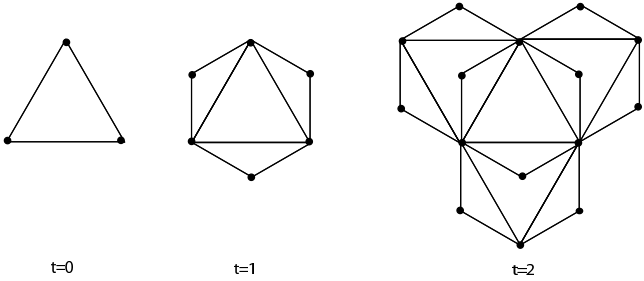}
\caption{The first three generations of the scale-free pseudo-fractal graph. At
each iteration step $t$, every edge generates an additional vertex, which is 
attached to the two vertices of this edge.}
\label{fig:fig1}
\end{figure} 

\section{Hierarchical Network}
The deterministic scale-free graph used in this work grows as follows: At 
each time step, every edge generates an additional vertex, which is attached 
to both end vertices of this edge. Initially, at $t=0$, we have a triangle 
of edges connecting three vertices, at $t=1$, the graph consists of 
$6$ vertices connected by $9$ edges, and so on (see Fig.\ref{fig:fig1}). The 
total number of vertices at iteration $t$ is 
\begin{equation}
N_{t}=\frac{3(3^{t}+1)}{2}
\label{eq:Nt}
\end{equation}
This simple rule produces a complex growing network. Such a graph is called 
a {\it pseudo-fractal}. In the next section, we present the use 
of this hierarchical network to find expressions that agree with
the simulated results of the Sznajd model on complex networks.  

\section{Renormalization Approach}
Our method can be very intuitive and is based on the method proposed 
by Galam to study bottom-up democratic voting by majority rule in a
square lattice \cite{galamI}, where the predictions of the results 
in all the lattice are based on the applications of the majority rule
over a basic cell of neighbors, called renormalization cell.

We find that given a neighborhood rule, it is enough to choose 
an appropriate generation of a hierarchical network for calculating
$P_{up}(p)|_{r,g}$, which agrees with the the numerical results 
of the model on growing networks The subscript index ${r,g}$ in
$P_{up}(p)|_{r,g}$ is to stress that the resulting function belongs 
to a chosen Sznajd rule ($r$) in a growing network ($g$). Subsequent 
self-iterations of $P_{up}(p)|_{r,g}$ result in a step function, 
i.e., $P^{n}_{up}(p)|_{r,g} = P_{up}(p)_{r,f}$, where the 
subscript index $f$ correponds to the result obtained for a network 
of fixed size. 
  
For a population fraction $p$ with opinion ``up'', the general method 
is as follows:
\begin{itemize}

\item Given a neighborhood rule $r$, the chosen basic cell 
corresponds to the minimum generation $t$ of the hierarchical network,
such that $r>N_{t}$ (the $r$ agents must have at least one agent to 
convince). We call this resulting number of nodes in the cell 
$n_{r}$.
  
\item  The probability of each possible configuration 
in a elementary cell is easily calculated, such that 
\begin{equation}
1=P_{all}(p)|_{r}=\sum_{k=2}^{n_{r}}B_{n_{r}k}p^{k}(1-p)^{n_{r}-k}.
\end{equation}
with the binomial coefficient $B_{n_{r}k}$ over the appropriate 
cell for the chosen rule:
\begin{equation}
B_{n_{r}k}=n_{r}!/[k!(n_{r}-k)!].
\end{equation}

\item  From all the configurations calculated above, we select
the subset that gives ``all up'' when applying the selected
Sznajd rule on the cell, the sum of all of them is $P_{up}(p)|_{r,g}$:
\begin{equation}
P_{up}(p)|_{r,g}= P_{all}(p)|_{r,up}
\end{equation} 
\end{itemize} 
Next, we illustrate the result  of the method with 
$r=2$ and $r=3$.
\begin{figure}[hbt]
\centering
\includegraphics[width=8.5cm]{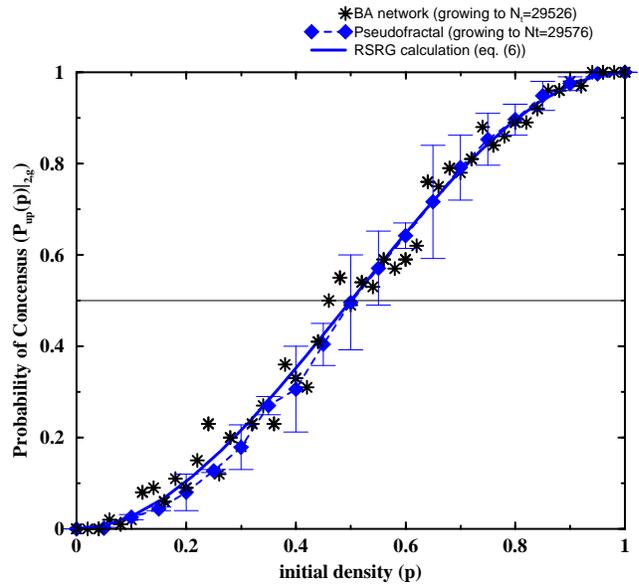}
\caption{Comparison between the function presented in Eq.~\ref{eq:gr2} (solid
  line) with Monte Carlo simulations on a growing pseudo-fractal (triangles
with error-bars) and on a growing $BA$ scale-free network (stars). In both
networks, $29576$ nodes are considered. We count the number of samples, out of
$1000$, for which the fixed point all ``up'' is obtained when different values 
for the initial concentration $p$ of nodes ``up'' are simulated for
rule $r=2$.}
\label{fig:fig2}
\end{figure} 

\subsection{Case $r=2$}
\subsubsection{Growing}

For $r=2$, the triangle of the generation $t=0$ is the basic cell. 
Thus $n_{r}=3$ and, for a given fraction $p$, all the possible
configurations are:
\begin{equation}
1=P_{all}(p)|_{2}=p^{3}+2p^{2}(1-p)+2p(1-p)^{2}+(1-p)^{3}
\end{equation}
If we apply the selected Sznajd rule $r=2$ over the triangle, only the 
configurations expressed in the first two terms of
the sum give ``all up''. Therefore: 
\begin{equation}
P_{up}(p)|_{2,g}=p^{3}+2p^{2}(1-p)=3{p}^{2}-2{p}^{3}
\label{eq:gr2}
\end{equation}
In Fig. \ref{fig:fig2}, we can see the good agreement of
Eq. \ref{eq:gr2} with the numerical results \cite{us} for the 
Sznajd model on a growing pseudo-fractal, as well as for the 
Barab\'asi-Albert scale-free network \cite{Barabasi:RMP}. 
\subsubsection{Fixed}
\begin{figure}[hbt]
\centering
\includegraphics[width=8.5cm]{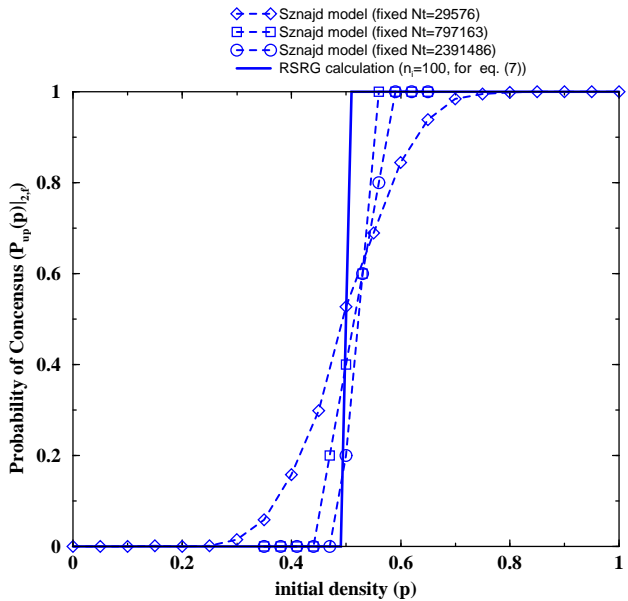}
\caption{Eq. \ref{eq:fr2} with $n_{i}=100$ (solid line) compared with 
simulations on a fixed pseudo-fractal networks with $N_{t}=29576$, 
$797163$ and $2391486$ nodes (dashed line with symbols). Other
simulation conditions as presented in the caption of Fig.~\ref{fig:fig2}.}
\label{fig:fig3}
\end{figure} 
In order to recover the reported result on a fixed network,
one makes renormalization iterations, which means
simply self-composing the Eq.~\ref{eq:gr2}:
\begin{equation}
P_{up}(p)|_{2,f}=P_{up}^{n_{i}}(p)|_{2,g},
\label{eq:fr2}
\end{equation}
and in the limit of large number of iterations
($n_{i}-1$), one recovers the step function observed 
numerically for the model on fixed networks. 
Note that the number of terms and the coefficients sizes increase very 
fast, as one can observe in the expression of only one 
composition:
\begin{equation}
P_{up}^{2}(p)|_{2,g}=27\,{p}^{4}-36\,{p}^{5}-42\,{p}^{6}+108\,{p}^{7}-72\,{p}^{
8}+16\,{p}^{9},
\label{eq:rf2}
\end{equation}
therefore, the multiple compositions presented in Fig.~\ref{fig:fig2} are
iterated with a computer. Figure \ref{fig:fig3} shows that the 
numerical simulations on large networks tend to the step function 
calculated from Eq. \ref{eq:rf2} with $n_{i}=100$.

\subsection{Case $r=3$}
\subsubsection{Growing}
\label{subsec:gr3}
\begin{figure}[hbt]
\centering
\includegraphics[width=8.5cm]{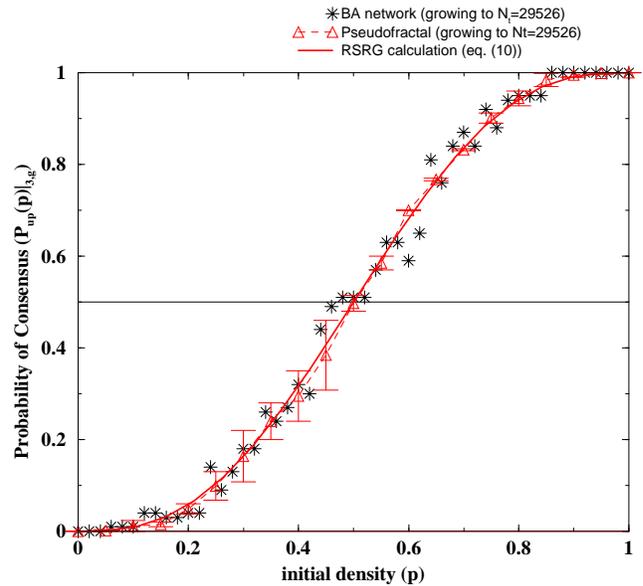}
\caption{Eq. \ref{eq:gr3} (solid line) compared with the results from
  the simulations on a growing pseudo-fractal (triangles with
  error-bars) and on a growing $BA$ scale-free network (stars) when 
$r=3$. The other simulation conditions are the same of Fig.~\ref{fig:fig2}.}
\label{fig:fig4}
\end{figure} 
The core of the method is the selection of the correct configurations 
after applying the Sznajd rule on it. As we will see for this rule, 
when the number of nodes in the renormalization cell is even,
there are some symmetrical configurations which can
have either ``all up'' or ``all down'' with
the same probability. In this case only half of them are summed 
to $P_{up}$. For $r=3$, the generation $t=1$ is the
basic cell. Thus $n_{r}=6$ and, for a given fraction $p$, all the 
possible configurations are:
\begin{equation}
1=P_{all}(p)|_{3}=(1+(1-p))^{6}.
\end{equation}
Note that the values of the binomial coefficient in the consecutive terms are:
$1,6,15,20,15,6,1$. 
From the $20$ configurations of the $4th$ term, there are $7$ that give ``all
up''(shown in Fig.~\ref{fig:fig6} at Appendix~\ref{app}), the corresponding $7$ opposed cases which give
``all down'', and $6$ symmetrical configurations shown in Fig.~\ref{fig:fig7} (Appendix~\ref{app})
that can give either ``all up'' or ``all down''. Therefore, these group 
of configurations contribute with $7+0.5 \times 6$, and we have:
\begin{equation}
P_{up}(p)|_{3,g}={p}^{6}+6\,{p}^{5}\left (1-p\right )+15\,{p}^{4}\left (1-p\right 
)^{2}+10\,{p}^{3}\left (1-p\right )^{3}
\label{eq:gr3}
\end{equation}

In Fig.~\ref{fig:fig4}, we see that Eq.~\ref{eq:gr3} agrees very well
with the numerical results \cite{us} for the Sznajd model on a growing
network when the rule $r=3$ is considered.

\subsubsection{Fixed}
\begin{figure}[hbt]
\centering
\includegraphics[width=8.5cm]{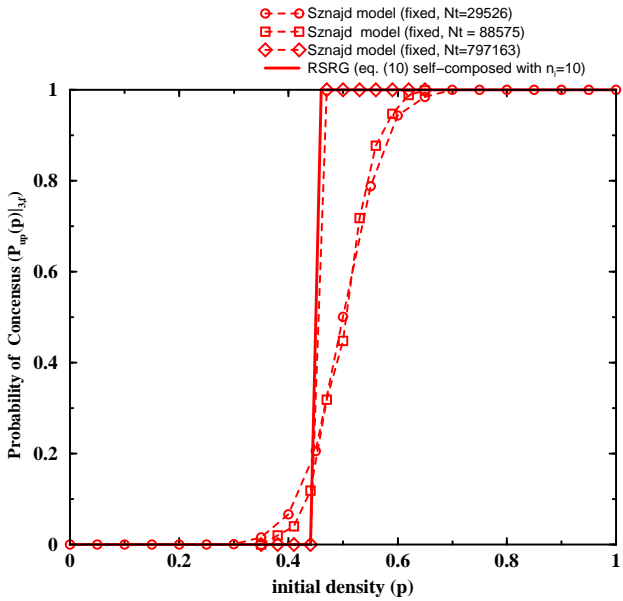}
\caption{Eq. \ref{eq:gr3} self-composed in $9$ steps 
(solid line) compared with the result from the simulations on  
a fixed pseudo-fractal networks with $r=3$ (dashed lines with
symbols). The other simulation conditions are the same as presented 
in the caption of Fig.~\ref{fig:fig3}.}
\label{fig:fig5}
\end{figure} 
The result of the composition for this case is far more complicated and
only $1$ self-composition  of eq.~\ref{eq:gr3} ($n_{i}=2$) already
needs a computer, as shows the following expression: 
\[
P_{up}(p)|_{3,g}^{2}=-1249989\,{p}^{12}+390897\,{p}^{11}-158184\,{p}^{10}+28561\,{p}^{9}-
\]
\[
643783179\,{p}^{18}+270741222\,{p}^{17}-100735317\,{p}^{16}+41109081\,
\]
\[
{p}^{15}-17504838\,{p}^{14}+5585931\,{p}^{13}-15244686567\,{p}^{24}+
\]
\[
11863411551\,{p}^{23}-7642674243\,{p}^{22}+4315583718\,{p}^{21}-
\]
\[
2347570026\,{p}^{20}+1281132990\,{p}^{19}-816731505\,{p}^{30}+
\]
\[
2281401855\,{p}^{29}-5100164190\,{p}^{28}+9199907505\,{p}^{27}-
\]
\[
13440029166\,{p}^{26}+15908268375\,{p}^{25}-2187\,{p}^{36}+65610\,{p}^{35}
\]
\[
-925101\,{p}^{34}+8148762\,{p}^{33}-50268195\,{p}^{32}+230706630\,
{p}^{31}
\]
In Fig.~\ref{fig:fig5} we see the step function obtained with only $9$ steps
of composition compared with the numerical results on a fixed network
of different sizes; as we see the results agree very well with the 
simulations of the model on large networks. 

\section{Conclusions}
Based on opinion formation rules of the usual Sznajd model, we use 
a renormalization approach to give an expression for the probability
of consensus into one opinion as a function of the initial fraction of
this opinion.

We show that for a given Sznajd rule it is enough to solve exactly the
model on an appropriate basic cell in order to find an expression
for the smooth function, found numerically for the model on
a growing network. Several self-compositions of the obtained
function give the step function observed for the model on a
network of fixed size. Further renormalization patterns
has to be tested, but in order to reproduce the results
of the Sznajd model on growing $SF$ networks, a
$SF$ hierarchical network must be chosen.     

The proposed method could be, in principle, extended to other 
types of neighborhood and more interestingly to many choices
of opinion ($q>2$), which is an feature of the model used
to simulate elections processes \cite{bernardes,us,slanina}, obtaining
results consistent with some empirical observations\cite{Costa:PRE}. 

\section*{Acknowledgments}
The authors would like to thank Dietrich Stauffer and 
Katarzyna Sznajd-Weron for useful discussions. 
MCG thanks Deutscher Aka\-demischer Austausch Dienst (DAAD), Germany,
for financial support.
 
\bigskip   
\bigskip

\appendix{{\bf Appendix A: Configurations with the same fraction of nodes ``up'' and ``down''}
\label{app}

Here we present some of the possible configurations applying the Sznajd rule
corresponding to the $r=3$ on its appropriate renormalization pattern ($n_{3}=6$).
In particular, we show the case of half of the nodes having opinion up, 
mentioned in Section \ref{subsec:gr3},and represented by the fourth
term in Eq.~\ref{eq:gr3}.
   
Figure \ref{fig:fig6} shows the $7$ configurations that give as a
result ``all up'', when applying the Sznajd rule, i.e., three
consecutive nodes with opinion $+1$ convince all their neighbors.
Note that interchanging $+$ and $-$, we have the $7$ configurations
for the opposed case of consensus ``all down''.
 
Figure~\ref{fig:fig7} presents the $6$ symmetrical configurations that have
$3$ consecutive nodes with $+1$, as well as $3$ nodes with $-1$ giving
consensus ``all up'' or ``all down'', respectively. Thus,
these configurations contribute wit $0.5 \times 6$ to the probability of 
consensus ``all up''($P_{p}$), as showed in Section~\ref{subsec:gr3}. 

\begin{figure}[hbt]
\centering
\includegraphics[width=8.5cm]{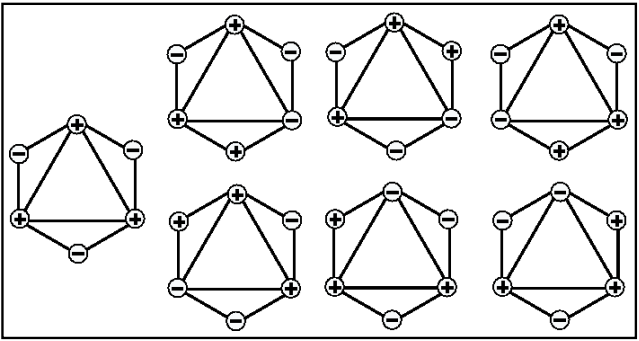}
\caption{Configurations that generate consensus ``all up'' with $r=3$ and the
same fraction of opinions ``up'' and ``down''.}
\label{fig:fig6}
\end{figure} 

\begin{figure}[hbt]
\centering
\includegraphics[width=8.5cm]{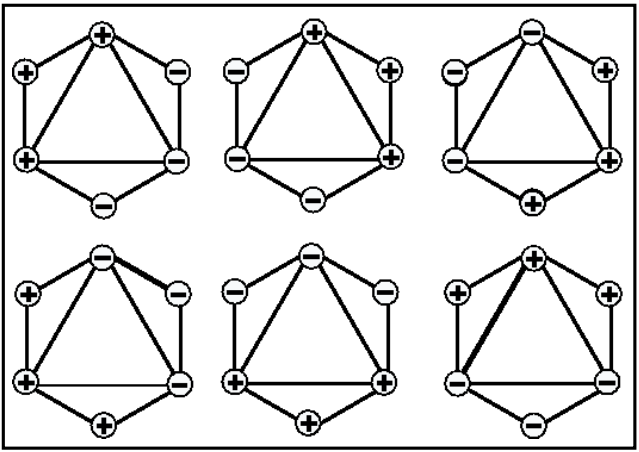}
\caption{Configurations that generate either consensus ``all  up'' or ``all down'' with $r=3$.}
\label{fig:fig7}
\end{figure} 



\end{document}